\documentstyle[a4,11pt,epsf]{article}
\epsfverbosetrue
\newcommand{\beq}{\begin{equation}}
\newcommand{\eeq}{\end{equation}}   

 0

\begin{document}

\begin{titlepage}
 
\begin{flushright}
Dublin Preprint: TCDMATH 97-06\\
Liverpool Preprint: LTH 399\\
cond-mat/9707039\\

\end{flushright}

\vspace{5mm}
 
\begin{center}
{\huge Fisher Zeroes and Singular Behaviour of the Two 
  \\ [3mm]
Dimensional Potts Model in the Thermodynamic Limit}\\[15mm]
{\bf R. Kenna\footnote{Supported by EU TMR
Project No. ERBFMBI-CT96-1757} 
\\
~\\ 
School of Mathematics, Trinity College Dublin,\\ Ireland\footnote{
Current address}\\
and\\
Department of Mathematical Sciences, Theoretical Physics Division\\
University of Liverpool, Liverpool L69 3BX, England
} \\[3mm]
July 1997
\end{center}
\begin{abstract}
The duality transformation is applied to the Fisher zeroes near 
the ferromagnetic critical point in the $q>4$ state two dimensional
Potts model. A requirement that the locus of the duals of the zeroes be
identical to the dual of the locus of zeroes in the thermodynamic limit 
(i) recovers the ratio of specific heat to internal
energy discontinuity at criticality and the
relationships between the discontinuities of higher cumulants and (ii) 
identifies duality with complex conjugation. Conjecturing that all 
zeroes governing ferromagnetic singular behaviour satisfy the latter 
requirement gives the full locus of such Fisher zeroes to be a circle.
This locus, together with the density of zeroes is then shown to be
sufficient to recover the singular part of the thermodynamic functions
in the thermodynamic limit.
\end{abstract}
\end{titlepage}

\newpage

\paragraph{Introduction}

The $q$--state Potts model \cite {Po52}, introduced in 1952 as a 
generalization of the Ising model \cite{Is25},
has become the generic model for the analytical 
and numerical study of first and second order phase transitions
\cite{Martinbook}.
Apart from the one dimensional case \cite{Po52}, 
the only solution which exists to date is for
the $q=2$ (Ising) model in two dimensions and in the absence of 
an external magnetic
field \cite{On44}.
The partition function for the standard Potts model is 
$ Z_L(\beta)  =  \sum_{\{\sigma_i\}} 
 \exp{(\beta \sum_{\langle ij \rangle}{\delta_{\sigma_i \sigma_j}})}
$
where $\beta = 1/(k_B T)$, and $T$ is the temperature. The spin 
$\sigma_i$ at site $i$ on a $d$--dimensional lattice takes values 
$1,2, \dots ,q$ and the total number of sites is $V=L^d$.
Despite the absence of a full solution for general $q$, some
exact and, indeed, rigorous results are obtainable in $d=2$.
The first of these is that, up to an irrelevant multiplicative
constant,  the form of the partition function is
unchanged under a duality transformation
in the thermodynamic limit \cite{Po52}.
In terms of the (low temperature expansion) variable $u=e^{-\beta}$,
this duality transformation is $u \rightarrow {\cal{D}}(u) $, where
\begin{equation}
 {\cal{D}}(u) = \frac{1-u}{1+(q-1)u}
\quad .
\label{duality}
\end{equation}
The critical temperature at which the phase transition occurs is
invariant under (\ref{duality}) and is given by \cite{Po52},
\begin{equation}
 {\cal{D}}(u_c) = u_c
 \quad {\rm{or}} \quad
  u_c = \frac{1}{1 + \sqrt{q}}
\quad .
\label{uc}
\end{equation}
Baxter has shown that {\em{at}} the critical point the model
is equivalent to a solvable homogeneous ice--type model 
\cite{Ba73,Ba83}.
By deriving the latent heat at criticality it was shown that the
phase transition in the two-dimensional model is first (second)
order for $q>4$ ($q \le 4$). In fact, for the $q>4$ case, the exact
values of the latent heat, the mean internal energy and the specific 
heat 
discontinuity (but not, for example, the mean specific heat) are known
\cite{Po52,Ba73,Ba83,Wu}.
The full form of the free energy (and derivable thermodynamic functions)
of the Potts model has, however, never been calculated for general $q$
and general $T$.
In the words of Baxter, solving the Potts model for general 
temperatures is, therefore, ``a very tantalising problem'' \cite{Ba83}.
The Potts model is reviewed in \cite{Wu}.

In this paper the problem is approached using a remarkably general
and recently derived result concerning the partition function zeroes
of models with a first order phase transition \cite{LeeI}.

For finite systems the zeroes of the partition function \cite{LY}
 are strictly complex (non--real).
As $L\rightarrow \infty$ one expects these zeroes to condense onto a
smooth
curve whose impact on to the real parameter axis precipitates the
phase transition.
Fisher \cite{Fi64} emphasized the application of zeroes in 
the  complex temperature plane to the study of temperature driven phase
transitions.
In particular, in \cite{Fi64}, the
Kaufman solution \cite{Ka49} of the two dimensional Ising model was used 
to
show that the Fisher zeroes (also called complex temperature zeroes
\cite{MaSh96}) are dense on two circles in the complex $u$--plane
in the thermodynamic limit.

Based on similarities with the Ising case, Martin and Maillard and Rammal 
\cite{Martin,MMR} conjectured that the locus of Fisher zeroes in the $d=2$, 
$q$--state Potts model be given by an extension of (\ref{uc}) to the complex
plane, namely ${\cal{D}}(u)=u^*$ where $u^*$ is the complex conjugate of
$u$ (although, on this basis alone `it is not clear where this requirement
comes from' \cite{Martin}). 
This identification yields a circle with centre $-1/(q-1)$ and
radius $\sqrt{q}/(q-1)$.
When $q=2$ this recovers the so--called
ferromagnetic Fisher circle of the Ising model \cite{Fi64}.
In the Ising case, the partition function
is actually a function of $u^2$. There, the second (so--called
antiferromagnetic) Fisher
circle comes from the map $u \rightarrow u^{-1}$
($\beta \rightarrow - \beta$).
Numerical investigations for small lattices at $q=3$ and $4$ \cite{Martin,MMR}
provided evidence that
the Fisher zeroes indeed lie on the circle given by  the identification
of duality with complex conjugation. However the numerics are highly
sensitive to the boundary conditions used and the situation far from
criticality remained unclear. Some progress was recently
made in the non--critical region using low temperature expansions
for $3\le q \le 8$ \cite{MaSh96}.

Recently, and on the basis of numerical results on small lattices 
(up to $L=7$) with
$q \le 10$, it has again been
conjectured that for finite lattices with self--dual boundary conditions,
and for other boundary conditions in the thermodynamic limit,
the zeroes in the ferromagnetic regime are on the above circle
\cite{ChHu96}. The conjecture of \cite{ChHu96} 
was, in fact, proven for infinite $q$ in \cite{WuRo96}.
This circle--conjecture is similar to another
recent conjecture \cite{AnMa94}, namely that the Fisher zeroes for
the $q$--state Potts model on a triangular lattice with pure three--site
interaction in the thermodynamic limit (which is also self--dual
\cite{Ba78}) lie on a circle and a segment of the negative real axis.

All of the above conjectures regarding the locus of Fisher zeroes
rely, at least in part, on numerical
approaches. 
In this paper, the problem is addressed analytically. A requirement
that taking the thermodynamic limit and application of the duality
transformation to the Fisher zeroes be commutative in the $q>4$ case
(i) recovers the ratio of specific heat discontinuity to latent heat
and corresponding relationships between the discontinuities 
of higher cumulants and (ii) analytically identifies duality with
complex conjugation.
Conjecturing that all 
zeroes governing ferromagnetic singular behaviour satisfy the latter 
requirement, the  locus of such Fisher zeroes is shown to indeed be 
a circle.
This locus, together with the density of zeroes is then shown to be
{\em{sufficient}} to recover the singular form of all thermodynamic 
functions
in the thermodynamic limit.

\paragraph{Thermodynamic Functions}
Consider, firstly, the finite--size system.
For finite $L$, the partition function  can be written as a 
polynomial of finite degree in $u$, and as such, can be expressed in
terms of its complex Fisher zeroes $u_j(L)$ \cite{LY} as
$ Z_L(\beta) \propto \prod_{j=1}^{dV}{(u-u_j(L))}$.
The free energy is defined by
$ \beta f(\beta) = - \ln{Z(\beta)}/V$.
The internal energy is therefore
\begin{equation}
 e(\beta)= \frac{\partial (\beta f)}{\partial \beta}
 =
 {\rm{cnst.}}
 + 
 \frac{u}{V}\sum_{j=1}^{dV}\frac{1}{u-u_j(L)}
\quad .
\label{e}
\end{equation}
The specific heat and the general $n^{\rm{th}}$ cumulant
are respectively defined as 
\begin{equation}
 c(\beta) =  -k_B \beta^2 \frac{\partial^2 (\beta f)}{\partial 
\beta^2}
\quad , \quad \quad 
 \gamma_n(\beta) = (-)^{n+1}\frac{\partial^n (\beta f)}{\partial 
\beta^n}
\quad .
\label{n}
\end{equation}
Using the notation
\begin{equation}
 \Delta \gamma_n
 \equiv
 \lim_{\beta \nearrow \beta_c}\gamma_n(\beta)
 -
 \lim_{\beta \searrow \beta_c}\gamma_n(\beta)
\quad,
\end{equation}
for the discontinuity in the $n^{\rm{th}}$ cumulant at the critical 
temperature, the exact results \cite{Po52,Ba73,Ba83,Wu}
(in the thermodynamic limit) are
\begin{eqnarray}
  {\overline{e}}
  & \equiv &
  \frac{1}{2}\left(\lim_{\beta \nearrow \beta_c}e(\beta) 
              +
              \lim_{\beta \searrow \beta_c}e(\beta)
             \right)
  =
  - \left( 1 + \frac{1}{\sqrt{q}} \right)
\quad ,
\label{ave}
 \\
  \Delta e
  & = &
   2 \left( 1 + \frac{1}{\sqrt{q}} \right)
   \tanh{\left( \frac{\Theta}{2} \right)}
   \prod_{n=1}^\infty{\tanh^2{(n\Theta)}}
\quad ,
\label{diffe}
 \\
  \Delta c
  & = &
   k_B \beta_c^2 \frac{\Delta e}{\sqrt{q}}
\quad ,
\label{diffc}
\end{eqnarray}
where $\Theta = \ln{\left(\sqrt{q/4}+\sqrt{q/4-1}\right)}$.
Further results include the general higher cumulant combination
$\gamma_n(\beta_c^-) - (-)^n\gamma_n(\beta_c^+)$
determinable from duality \cite{Wu,BhLa97,JaKa97}.

\paragraph{Partition Function Zeroes}
Recently, Lee \cite{LeeI} has derived a general theorem for first
order phase transitions in which the partition function zeroes
can be  expressed in terms of the discontinuities in the thermodynamic 
functions (for finite size as well as in the infinite volume limit).
For a system with a temperature--driven phase transition, Lee's result
for the Fisher zeroes is
\begin{equation}
 \frac{\ln{q}}{V \Delta e}
 \pm
 i\frac{(2j-1)\pi}{V \Delta e}
 =
 \beta_c t + 
 \frac{t^2}{2!}\frac{\Delta c/k_B}{\Delta e} 
 + \sum_{n=3}^\infty{
 \frac{(\beta_c t)^n}{n!} 
 \frac{\Delta \gamma_n}{\Delta e} 
 }
\label{core}
\end{equation}
where the reduced temperature is $t=1-\beta/\beta_c$. Inverting, we find
\cite{LeeI}
\begin{eqnarray}
 \beta_c {\rm{Re}} t_j(L)
 & = &
 A_1 I_j^2 + A_3 I_j^4  + A_5  I_j^6
 + \dots
 + {\cal{O}}(1/V)
\quad ,
\nonumber \\
 \pm \beta_c {\rm{Im}}t_j(L) 
 & = &
 I_j
 +
 A_2 I_j^3
 +
 A_4 I_j^5
 + \dots
 + {\cal{O}}(1/V)
\label{Itj}
\quad ,
\end{eqnarray}
where $ I_j = (2j-1)\pi/(V\Delta e)$
and  ${\cal{O}}(1/V)$ represents terms which vanish in the
infinite volume limit and where the coefficients $A_n$ are easily
calculable, the first few being \cite{LeeI}
\begin{eqnarray}
 A_1
 & = & 
 \frac{\Delta c / k_B \beta_c^2}{2 \Delta e}
\quad ,
\label{aa1} \\
 A_2  & =  & 
 - 2A_1^2
 + \frac{\Delta \gamma_3}{3!\Delta e}
\quad ,
\label{aa2} \\
 A_3  & =  & 
 - 5A_1^3
 + 5A_1
    \frac{\Delta \gamma_3}{3!\Delta e}
 - \frac{\Delta \gamma_4}{4!\Delta e}
\quad ,
\label{aa3} \\
 A_4  & =  & 
 14 A_1^4 - 21 A_1^2 \frac{\Delta \gamma_3}{3!\Delta e}
 + 3 \left( \frac{\Delta \gamma_3}{3!\Delta e} \right)^2
 + 6 A_1  \frac{\Delta \gamma_4}{4!\Delta e} 
 - \frac{\Delta \gamma_5}{5!\Delta e}
\quad ,
\label{aa4} \\
 A_5  & =  & 
 42 A_1^5 - 84 A_1^3 \frac{\Delta \gamma_3}{3!\Delta e}
 + 28 A_1 \left( \frac{\Delta \gamma_3}{3!\Delta e} \right)^2
 + 28 A_1^2  \frac{\Delta \gamma_4}{4!\Delta e}
  - 7  \frac{\Delta \gamma_3}{3!\Delta e} 
      \frac{\Delta \gamma_4}{4!\Delta e} 
 - 7 A_1 \frac{\Delta \gamma_5}{5!\Delta e}
 + \frac{\Delta \gamma_6}{6!\Delta e}
\quad 
\label{aa5}
\end{eqnarray}

\paragraph{The Locus of Zeroes}
From (\ref{Itj}) the real part of the 
zeroes (in the thermodynamic limit) can be expressed
in terms of their imaginary parts as
$ \beta_c {\rm{Re}}t = {\cal{L}}(\beta_c {\rm{Im}}t) $
where
\begin{equation}
 {\cal{L}}(\theta)
 = 
 A_1\theta^2
 +
 (-2A_1A_2+A_3)\theta^4
 +
 (7A_1A_2^2-2A_1A_4-4A_2A_3+A_5)\theta^6
 + \dots\quad .
\label{locus}
\end{equation}
The zeroes are thus seen to lie on a curve. In the complex $u$ upper 
half--plane the equation of this curve is 
\begin{equation}
 \gamma^{(+)}(\theta) = u_c e^{{\cal{L}}(\theta)+i\theta}
\quad .
\end{equation}
This defines the locus of zeroes in the infinite volume limit.

\paragraph{The Dual of the Locus of Zeroes}
Applying the duality transformation (\ref{duality}) to
$\gamma^{(+)}(\theta)$ and expanding in powers of $\theta$ gives
\begin{eqnarray}
 \lefteqn{
{\rm{Re}}  {\cal{D}} \left(   {\gamma^{(+)}}(\theta) \right)
 =
u_c
\Big[
      1 +
      \frac{\theta^2}{2q}\left(2\sqrt{q}-2A_1q -q \right) 
      -\frac{\theta^4}{24q^2}
      \left(
            24\sqrt{q}-36q+14q^\frac{3}{2}-q^2
      \right.
}
\nonumber \\
 &  &
 \left.
     - 72A_1q
     +72A_1q^\frac{3}{2} 
     -12A_1q^2+24A_1^2q^\frac{3}{2}
     -12A_1^2q^2-48A_1A_2q^2
     +24A_3q^2   
 \right)
  + \dots    
\Big]
\quad ,
\label{rexpd}
\\
 \lefteqn{
 {\rm{Im}}  {\cal{D}}\left(   {\gamma^{(+)}}(\theta) \right)
= -u_c 
\Big[
      \theta 
      - \frac{\theta^3}{6q}
        \left( 6-6q^\frac{1}{2}+q-12A_1q^\frac{1}{2}+6A_1q \right)
    -\frac{\theta^5}{120q^2}
\left(
      -120
      +240q^\frac{1}{2}
\right.
}
\nonumber \\
 & & 
\left.
      -150q
      +30q^\frac{3}{2}
      -q^2
      +480A_1q^\frac{1}{2}
      -720A_1q
      +280A_1q^\frac{3}{2}
      -20A_1q^2
      -360A_1^2q
      +360A_1^2q^\frac{3}{2}
\right.
\nonumber \\
 & & 
\left.
      -60A_1^2q^2
      +480A_1A_2q^\frac{3}{2}
      -240A_1A_2q^2
      -240A_3q^\frac{3}{2}
      +120A_3q^2
      \right)
+ \dots
\Big] 
\quad .
\label{iexpd}
\end{eqnarray}

\paragraph{The Locus of the Duals of the Zeroes}

Alternatively, 
applying the duality transformation (\ref{duality}) directly
to the $j^{\rm{th}}$ zero  in the finite--size system 
and expanding again, gives
\begin{eqnarray}
 \beta_c {\rm{Re}} t_j^D(L)
 & = &
 A_1^D
 I_j^2
 + A_3^D
 I_j^4
 + A_5^D
 I_j^6
 + \dots
 + {\cal{O}}(1/V)
\quad ,
\nonumber
\\
 \mp \beta_c {\rm{Im}}t_j^D (L)
 & = &
 I_j
 +
 A_2^D 
 I_j^3
 +
 A_4^D
 I_j^5
 + \dots
 + {\cal{O}}(1/V)
\quad ,
\label{ItjD}
\end{eqnarray}
where the first few coefficients $A_n^D$ are
\begin{eqnarray}
 A_1^D
 & = & 
 q^{-\frac{1}{2}}-A_1
\quad ,
\label{ad1}
\\
 A_2^D  & =  &
 -q^{-1}+2 q^{-\frac{1}{2}}A_1 + A_2
\quad ,
\label{ad2}
\\
 A_3^D  & =  &
 -\frac{1}{12} q^{-\frac{1}{2}}
 - q^{-\frac{3}{2}}
 + 3 q^{-1} A_1
 - q^{-\frac{1}{2}}A_1^2
 + 2 q^{-\frac{1}{2}} A_2 
 - A_3
\label{ad3}
\\
 A_4^D & = &
 \frac{1}{4} q^{-1} 
 + q^{-2}
 - \frac{1}{3} q^{-\frac{1}{2}} A_1
 - 4 q^{-\frac{3}{2}} A_1
 + 3 q^{-1} A_1^2
 - 3 q^{-1} A_2
 + 2 q^{-\frac{1}{2}} A_1 A_2
 +2 q^{-\frac{1}{2}} A_3
 + A_4
\label{ad4}
\\
 A_5^D  & =  &
 \frac{1}{360} q^{-\frac{1}{2}}
 + \frac{1}{2} q^{-\frac{3}{2}}
 + q^{-\frac{5}{2}}
 - \frac{5}{4} q^{-1}A_1
 - 5 q^{-2}A_1
 + 6 q^{-\frac{3}{2}}A_1^2
 + \frac{1}{2} q^{-\frac{1}{2}}A_1^2
 - q^{-1}A_1^3
 - \frac{1}{3} q^{-\frac{1}{2}}A_2
\nonumber \\
 & & 
 - 4 q^{-\frac{3}{2}}A_2
 + 6 q^{-1}A_1 A_2
 + q^{-\frac{1}{2}}A_2^2
 + 3 q^{-1} A_3 
 - 2 q^{-\frac{1}{2}}A_1 A_3
 + 2 q^{-\frac{1}{2}}A_4
 - A_5
\quad .
\label{ad5}
\end{eqnarray}
From (\ref{ItjD}) the real part of the dual
zeroes can be expressed in terms of their imaginary parts 
in the thermodynamic limit as
$ \beta_c {\rm{Re}}t_j^D
   = {\cal{L}}^D({\rm{Im}}t_j^D) $
where
\begin{equation}
 {\cal{L}}^D(\theta)
 = 
 A_1^D\theta^2
 +
 (-2A_1^DA_2^D+A_3^D)\theta^4
 +
 (7A_1^D{A_2^D}^2-2A_1^DA_4^D-4A_2^DA_3^D-A_5^D)\theta^6
 + \dots
\quad .
\label{locusD}
\end{equation}
Therefore, the locus of the dual of the upper half--plane zeroes 
in the thermodynamic limit is given by
\begin{equation}
 {\gamma^{(+)}}^D(\theta) = u_c e^{{{\cal{L}}^D}(\theta)-i\theta}
\quad .
\end{equation}
The expansion of this locus of duals is
\begin{eqnarray}
 \lefteqn{
{\rm{Re}}{\gamma^{(+)}}^D(\theta)
}
\nonumber \\
 & = &
 u_c \Big[
           1 +
           \frac{\theta^2}{2!}\left(-1+2A_1^D\right) +
           \frac{\theta^4}{4!}\left(1-12A_1^D-48A_1^DA_2^D+24A_3^D
                          +12{A_1^D}^2\right)
     + \dots
     \Big]
\quad ,
\label{rexpld}
\\
 \lefteqn{
 {\rm{Im}}{\gamma^{(+)}}^D(\theta)
}
\nonumber \\
 & = &
 -u_c \Big[\theta 
           + \frac{\theta^3}{3!}\left(-1+6A_1^D \right)
           + \frac{\theta^5}{5!}
             \left(1-20A_1^D+60{A_1^D}^2-240A_1^DA_2^D+120A_3^D\right)
           + \dots
     \Big]
\quad .
\label{iexpld}
\end{eqnarray}
\paragraph{Identification of the Dual of the Locus with the Locus of 
the duals}

In deriving the dual of the locus of zeroes (\ref{rexpd}) and 
(\ref{iexpd}),
the duality transformation was applied {\em{after}} the thermodynamic
limit of the positions of the zeroes (i.e., their locus) was taken.
In (\ref{rexpld}) and (\ref{iexpld}) the duality transformation was
applied to the zeroes {\em{before}} taking the thermodynamic limit.
Even in the case where the finite--$L$ system does not have
duality--preserving
boundary conditions, taking the thermodynamic limit restores 
self--duality.
The dual of the (thermodynamic limit) locus of zeroes and the 
(thermodynamic limit) locus of the duals of the zeroes
must be identical. We demand, therefore, that
\begin{equation}
  {\cal{D}}\left( \gamma^{(+)} \right)
  \equiv
  {\gamma^{(+)}}^D
\label{demand}
\end{equation}
order by order in the expansion in $\theta$.
Up to ${\cal{O}}(\theta^2)$ this is trivial.
To ${\cal{O}}(\theta^3)$ and (separately at) 
${\cal{O}}(\theta^4)$ they are identical if
$
 A_1=1/(2\sqrt{q})
$.
From (\ref{aa1}), this is the result (\ref{diffc}).
The identity (\ref{demand}) at ${\cal{O}}(\theta^5)$ and (separately at)
${\cal{O}}(\theta^6)$ gives
$A_3 = A_2/\sqrt{q} - q^{-3/2}(q-3)/24$, which from 
(\ref{aa2}) and (\ref{aa3}) means that 
\begin{equation}
 \Delta \gamma_4 = \frac{6}{\sqrt{q}}\Delta \gamma_3
 + \frac{q-6}{q^{3/2}}\Delta e 
\quad .
\label{a4}
\end{equation}
Higher order results are obtainable using a computer algebra system
such as Maple. 
To orders ${\cal{O}}(\theta^7)$ and ${\cal{O}}(\theta^{8})$
and (separately) 
to orders ${\cal{O}}(\theta^9)$ and ${\cal{O}}(\theta^{10})$
one finds
\begin{equation}
 \frac{\Delta \gamma_6}{6! \Delta e}
  = 
 \frac{5}{2q^{1/2}}  \frac{\Delta \gamma_5}{5! \Delta e}
 + \frac{q-20}{8q^{3/2}} \frac{\Delta \gamma_3}{3! \Delta e}
 + \frac{1}{6! q^{1/2}} 
 - \frac{1}{8q^{3/2}} 
 + \frac{1}{2q^{5/2}} 
\quad ,
\label{a6}
\end{equation}
and
\begin{eqnarray}
 \frac{\Delta \gamma_8}{8! \Delta e}
 & = & 
 \frac{7}{2 q^{1/2}}  \frac{\Delta \gamma_7}{7! \Delta e}
 + \left( \frac{5}{24 q^1/2} - \frac{35}{4 q^{3/2}}
   \right)
 \frac{\Delta \gamma_5}{5! \Delta e}
 + \left( \frac{3}{6!q^{1/2}} - \frac{15}{16 q^{3/2}}
 +\frac{21}{2q^{5/2}}
   \right)
 \frac{\Delta \gamma_3}{3! \Delta e}
\nonumber \\
 & &
 + \frac{1}{8! q^{1/2}} 
 - \frac{23}{960q^{3/2}} 
 + \frac{5}{8q^{5/2}} 
 - \frac{17}{8q^{7/2}} 
\quad ,
\label{a78}
\end{eqnarray}
respectively.
These results and further results for the higher cumulants at 
criticality are also obtainable directly from the duality transformation
(\ref{a4}) (see \cite{Wu,BhLa97,JaKa97}).

\paragraph{The Full Ferromagnetic Locus of Zeroes}

Putting the above equations into
(\ref{ad1}) -- (\ref{ad5}) (and their higher order equivalents)
yields $A_j^D  =  A_j$ (this has been verified up to $j=8$).
Therefore (at least up to $\theta^{10}$) the dual of the locus of zeroes
is the complex conjugate of the original locus of zeroes.
We now assume that this is the case for all $\theta$. Then, the full
ferromagnetic locus of zeroes 
(that part of the full locus which intersects the real temperature
axis at the physical ferromagnetic critical point) is found by 
identifying \cite{Martin,MMR}
\begin{equation}
 {\cal{D}}\left({\gamma^{(+)}}(\theta)\right) =  
{\gamma^{(+)}}^*(\theta)
\quad ,
\label{fflz}
\end{equation}
where  ${\gamma^{(+)}}^*$ represents the complex conjugate of
$\gamma^{(+)}$.
The full ferromagnetic locus is then \cite{Martin,MMR}
\begin{equation}
 \gamma(\theta)=\frac{1}{q-1}\left(-1+\sqrt{q}e^{i\theta}\right)
\quad .
\label{ff}
\end{equation}

This circular locus is analogous to the circle theorem of Lee
and Yang \cite{LY}. In the field driven case, where one is
interested in Lee--Yang zeroes in the complex $z=\exp{h}$
plane ($h$ is an external magnetic field), formulae
analogous to (\ref{Itj}) -- (\ref{aa5}) apply where $e$, $c$,
etc. are replaced by the corresponding derivative of the
free energy with respect to $h$ (the magnetization $m$,
the susceptibility $\chi$, etc.). There, the partition function
is unchanged under $h \rightarrow -h$ and consequently
$\Delta \gamma_l =0$ for even $l$. Therfore ${\cal{L}}(\theta)=0$ and
the locus of zeroes is $z=\exp{i \theta}$. This is Lee's proof of
the Lee--Yang theorem \cite{LeeI}. One observes that considering 
$h \rightarrow -h$ as a self--duality map and identifying it with
complex conjugation yields this locus.

\paragraph{The Singular Parts of the Thermodynamic Functions in the
Thermodynamic Limit}

From (\ref{core}), the density of zeroes in the temperature driven
case is \cite{Sa94}
\begin{eqnarray}
 \lefteqn{
 g(\theta)=\lim_{V\rightarrow \infty}{\frac{1}{V} \frac{dj}{d\theta} }= 
 \frac{\Delta e}{2\pi}\left(1+\frac{1}{(q-1)\gamma(\theta)}\right)
 \left\{
        1 
        + 
        \frac{\Delta c /k_B \beta_c^2}{\Delta e}
        \ln{ \left(
           (\sqrt{q}+1)\gamma(\theta)
             \right)
           }
 \right.
}
\nonumber \\
 &  & 
\quad \quad  \left.
         +
         \frac{1}{2!}
         \frac{\Delta\gamma_3}{\Delta e}
         \left(
        \ln{ \left(
           (\sqrt{q}+1)\gamma(\theta)
             \right)
           }
         \right)^2
         +
         \frac{1}{3!}
         \frac{\Delta\gamma_4}{\Delta e}
         \left(
        \ln{ \left(
           (\sqrt{q}+1)\gamma(\theta)
             \right)
           }
         \right)^3
   + \dots
 \right\}
\quad .
\label{density}
\end{eqnarray}
The internal energy is (from (\ref{e}) or \cite{AbeSuzuki,rak})
\begin{equation}
 e = {\rm{cnst.}}
     +
     u
     \int_0^{2\pi}{
                  \frac{g(\theta)}{u-\gamma(\theta)}d\theta
                  }
\quad .
\label{egl}
\end{equation}
Therefore, from (\ref{ff}), (\ref{density}) and (\ref{egl}), 
the internal energy is
\begin{eqnarray}
 e(\beta < \beta_c) 
 & = & e_0 \quad \\
 e(\beta > \beta_c) 
 & = &
 e_0 
 -
 \Delta e
 \left\{
        1 + \frac{\Delta c/k_B \beta_c^2}{\Delta e}
            (\beta_c - \beta)
          + \frac{1}{2!}
            \frac{\Delta \gamma_3}{\Delta e}
            (\beta_c - \beta)^2
          + \frac{1}{3!}
            \frac{\Delta \gamma_4}{\Delta e}
            (\beta_c - \beta)^3
          + \dots
 \right\}
\quad ,
\label{e2}
\end{eqnarray}
with $e_0$ a constant
(one expects that when separate Fisher loci which don't cross the
positive real temperature axis are accounted for, $e_0$ becomes
temperature dependent).
At $\beta_c$ the internal energy discontinuity
$e(\beta=\beta_c^-)- e(\beta=\beta_c^+) = \Delta e$ is recovered.
Appropriate differentiation recovers the discontinuities in specific
heat and higher cumulants.
Thus the full locus (\ref{ff}) and the density (\ref{density})
are sufficient to give the {\em{singular}} parts of the thermodynamic
functions in the infinite volume limit.

\paragraph{Conclusions}

In summary,
we have applied the duality transformation (\ref{duality}), under which 
the $d=2$ $q$--state Potts model is invariant, to the Fisher
zeroes recently found by Lee \cite{LeeI} for systems with a first order
phase transition. The requirement that the dual of the locus of zeroes
be identical to the locus of the duals of the zeroes 
in the thermodynamic limit (i) recovers the ratio of specific 
heat to internal energy discontinuity
at criticality and the relations between the 
discontinuities of higher cumulants and (ii) identifies duality with
complex conjugation. 

Conjecturing that all zeroes
governing ferromagnetic critical behaviour satisfy (ii) gives that
this full locus is the circle (\ref{ff}) in the complex $u$--plane.
The equation (\ref{ff}) was first conjectured by Martin and Maillard
and Rammal \cite{Martin,MMR} on the basis of analogous Ising results 
\cite{Fi64}. The
same conjecture, based on numerical results for small lattices 
was made in \cite{ChHu96} and proven for infinite
$q$ in \cite{WuRo96} (see also \cite{AnMa94}).

The locus (\ref{ff}), together with the density of zeroes is  
sufficient to recover the {\em{singular}} parts of all thermodynamic 
functions in the thermodynamic limit.
It is to be expected that the {\em{regular}} parts come from 
separate loci of zeroes which don't cross the positive
real temperature axis. 

\paragraph{Acknowledgements}
The author wishes to thank Alan Irving, Wolfhard Janke and Jim Sexton
for stimulating discussions.

\newpage
 



\begin{thebibliography}{1234567} 
\newcommand{\bibi}[1]{\bibitem{#1}}
\newcommand{\authors}[1]{#1, }
\newcommand{\journal}[1]{#1}
\newcommand{\volume}[1]{{\bf #1}}
\newcommand{\myyear}[1]{(#1)}
\newcommand{\page}[1]{#1}
\newcommand{\mytitle}[1]{}
\newcommand{\keywords}[1]{}
\newcommand{\kw}[1]{}
\bibitem{Po52}
 R.B. Potts,
 Proc. Camb. Phil. Soc. {\bf{48}} (1952) 106.
\bibitem{Is25}
 E. Ising, Z. Physik {\bf{31}} (1925) 253.
\bibitem{Martinbook}
 P.P. Martin, {\em{Potts Models and Related Problems in Statistical 
                        Mechanics}} (World Scientific, Singapore, 1991).
\bibitem{On44}
 L. Onsager, Phys. Rev. {\bf{65}} (1944) 117.
\bibitem{Ba73}
 R.J. Baxter, J. Phys. C {\bf{6}} (1973) L445;
              J. Stat. Phys. {\bf{28}} (1982) 1.
\bibitem{Ba83}
R.J. Baxter,
Exactly Solved Models in Statistical Physics,
(Acadamic Press, London, 1982).
\bibitem{Wu}
 F.Y. Wu,
 Rev. Mod. Phys.
{\bf{54}} (1982)  235.
\bibitem{LeeI}
K.-C. Lee
Phys. Rev. Lett.
{\bf{73}} (1994)  2801.
\bibitem{LY}
C.N. Yang and T.D. Lee
Phys. Rev.
{\bf{87}} (1952)  404; ibid. 410.
\bibitem{Fi64}
M.E. Fisher,
in {\em Lectures in Theoretical Phys\-ics, Vol. VIIC},
ed. W.E. Brittin, (Gordon and Breach, New York: 1968) 1.
\bibitem{Ka49}
B. Kaufman,
Phys. Rev. {\bf{76}} (1949) 1232.
\bibitem{MaSh96}
 V. Matveev and R. Shrock,
 Phys. Rev. E {\bf{54}} (1996) 6174.
\bibitem{Martin}
 P.P. Martin, Nucl. Phys. B {\bf{225}} (1983)  497.
\bibitem{MMR}
 J.M. Maillard and R. Rammal, J. Phys. A {\bf{16}} (1983) 353;
 P.P. Martin, {\em{Integrable Systems in Statistical Mechanics}},
 ed. G.M. d'Ariano, A. Montorsi and M.G. Rasetti,
 (Singapore: World Scientific, 1985), 129;
 P.P. Martin and J.M. Maillard, J. Phys. A {\bf{19}} (1986) L547;
 ibid. {\bf{20}} (1987) L601.
\bibitem{ChHu96}
C.-N. Chen, C.-K. Hu and  F.Y. Wu,
 Phys. Rev. Lett., {\bf{76}} (1996) 169.
\bibitem{WuRo96}
 F.Y. Wu, G. Rollet, H.Y. Huang, J.M. Maillard, C.-K. Hu and C.-N. Chen,
 Phys. Rev. Lett., {\bf{76}} (1996) 173.
\bibitem{AnMa94}
     J.C. Angl{\`e}s d'Auriac, J. M. Maillard, G. Rollet and F.Y. Wu
                                            Physica A 206 (1994) 441.
\bibitem{Ba78}
 R.J. Baxter, H.N.V. Temperley and S. Ashley
              Proc. Roy. Soc. London, Ser. A {\bf{358}} (1978) 535.
\bibitem{BhLa97}
T. Bhattacharya, R. Lacaze and A. Morel, 
Nucl. Phys. B {\bf{435}} (1995) 526;
J. Phys. I France {\bf{7}} (1997) 81.
\bibitem{JaKa97}
 W. Janke and S. Kappler, 
Phys. Lett. A {\bf{197}} (1995) 227;
J. Phys. I France {\bf{7}} (1997) 663.
\bibitem{AbeSuzuki}
R. Abe, Prog. Theor. Phys. {\bf 37}, (1967) 1070;
ibid. {\bf{38}} 322;
M. Suzuki, Prog. Theor. Phys. {\bf 38}, (1967) 1243.
\bibitem{rak}    R. Kenna and C.B. Lang, Phys. Rev. E49 (1994) 5012;
 R. Kenna and A.C. Irving, Nucl. Phys. B 485 (1997) 583.
\bibitem{Binder}
 K. Binder, Z. Phys. B {\bf{43}} (1981) 119.          
\bibitem{Sa94}
The existence of a unique density of zeroes in the thermodynamic limit 
was proven in M. Salmhofer,
Nucl. Phys. B, Proc. Suppl. {\bf{30}} (1993) 81;
Helv. Phys. Acta {\bf{67}} (1994) 257.




\end{thebibliography}
\end{document}